\documentclass{Interspeech}

\usepackage{graphicx}
\usepackage{pifont}

\usepackage{amsmath}
\usepackage{amssymb}

\interspeechcameraready

\title{MM-MovieDubber: Towards Multi-Modal Learning for Multi-Modal \\Movie Dubbing}

\author[affiliation={1}]{Junjie}{Zheng}
\author[affiliation={1}]{Zihao}{Chen}
\author[affiliation={1}]{Chaofan}{Ding}
\author[affiliation={1}]{Yunming}{Liang}
\author[affiliation={2}]{Yihan}{Fan}
\author[affiliation={1}]{\\Huan}{Yang}
\author[affiliation={3}]{Lei}{Xie}
\author[affiliation={1}]{and Xinhan}{Di}

\affiliation{AI Lab}{Giant Network}{China}
\affiliation{}{The East China University of Science and Technology}{China}
\affiliation{ASLP@NPU}{Northwestern Polytechnical University}{China}
\email{*@ztgame.com, Y80220125@mail.ecust.edu.cn, lxie@nwpu.edu.cn}


\usepackage{comment}

\begin{document}

\maketitle


\begin{abstract}
Current movie dubbing technology can produce the desired speech using a reference voice and input video, maintaining perfect synchronization with the visuals while effectively conveying the intended emotions. However, crucial aspects of movie dubbing, including adaptation to various dubbing styles, effective handling of dialogue, narration, and monologues, as well as consideration of subtle details such as speaker age and gender, remain insufficiently explored. To tackle these challenges, we introduce a multi-modal generative framework. First, it utilizes a multi-modal large vision-language model (VLM) to analyze visual inputs, enabling the recognition of dubbing types and fine-grained attributes. Second, it produces high-quality dubbing using large speech generation models, guided by multi-modal inputs. Additionally, a movie dubbing dataset with annotations for dubbing types and subtle details is constructed to enhance movie understanding and improve dubbing quality for the proposed multi-modal framework. Experimental results across multiple benchmark datasets show superior performance compared to state-of-the-art (SOTA) methods.
In details, the LSE-D, SPK-SIM, EMO-SIM, and MCD exhibit improvements of up to 1.09\%, 8.80\%, 19.08\%, and 18.74\%, respectively.

\end{abstract}

\section{Introduction}

Dubbing involves adding the correct human voice to a video's dialogue, ensuring synchronization with the characters' lip movements, and conveying the emotions of the scene. Existing dubbing methods can be categorized into two groups, each focusing on learning different styles of key prior information to generate high-quality voices. The first group focuses on learning effective speaker style representations \cite{chen2022v2c,hassid2022more,wan2018generalized,cong2023learning}. The second group aims to learn appropriate prosody by utilizing visual information from the given video input \cite{hu2021neural,lee2023imaginary,zhao2024mcdubber}. However, the accuracy of these priors is often insufficient due to the limitations of various modules, which impacts the overall quality of speech generation.
Additionally, with the rapid advancement of large language models (LLMs) and related methods, multi-modal large vision language models (VLMs) have increasingly demonstrated their potential in multi-modal understanding tasks. These models can integrate information from various modalities for effective understanding, such as audio-text \cite{penamakuri2024AudiopediaAudioQA} and vision-text \cite{Insight, cheng2024videollama2advancingspatialtemporal}. These advancements in VLMs understanding capabilities hold promise for accurately providing dubbing types and fine-grained attributes.

\begin{figure}[t!]
    \centering
    \includegraphics[scale=0.60]{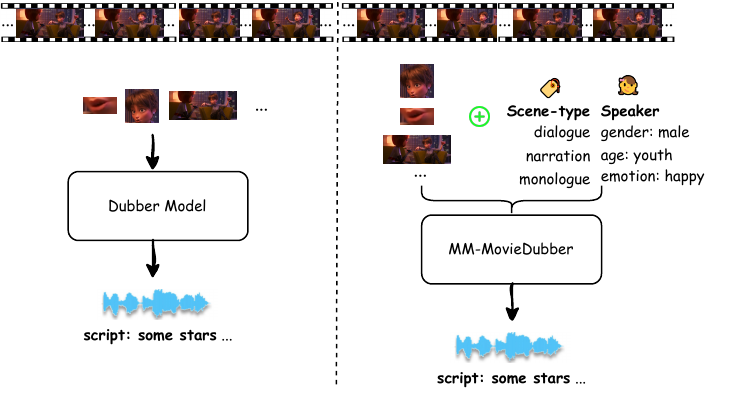}
    \caption{Curent Dubbing models \cite{congLearningDubMovies2023,cong2024styledubbermultiscalestylelearning,zhang2024from} (Left). Proposed Dubbing Models (Right) For dubbing types and fine-grained attributes.}
    \label{fig:model_pipeline}
\end{figure}

Therefore, we introduce a multi-modal generative framework for movie dubbing that incorporates an understanding of dubbing types and fine-grained attributes. First, a multi-modal large vision-language model (VLM) is trained to improve the understanding of dubbing types and fine-grained attributes from video inputs. Second, a large speech generation model is trained with designed control mechanisms guided by multi-modal conditions. Finally, we construct a movie dubbing dataset with diverse annotations, including dubbing types(Dialogue, Narration and Monologue) and fine-grained attributes, to improve the movie understanding and dubbing quality of the proposed model.  

\begin{figure*}[h]
    \centering
    \includegraphics[scale=0.60]{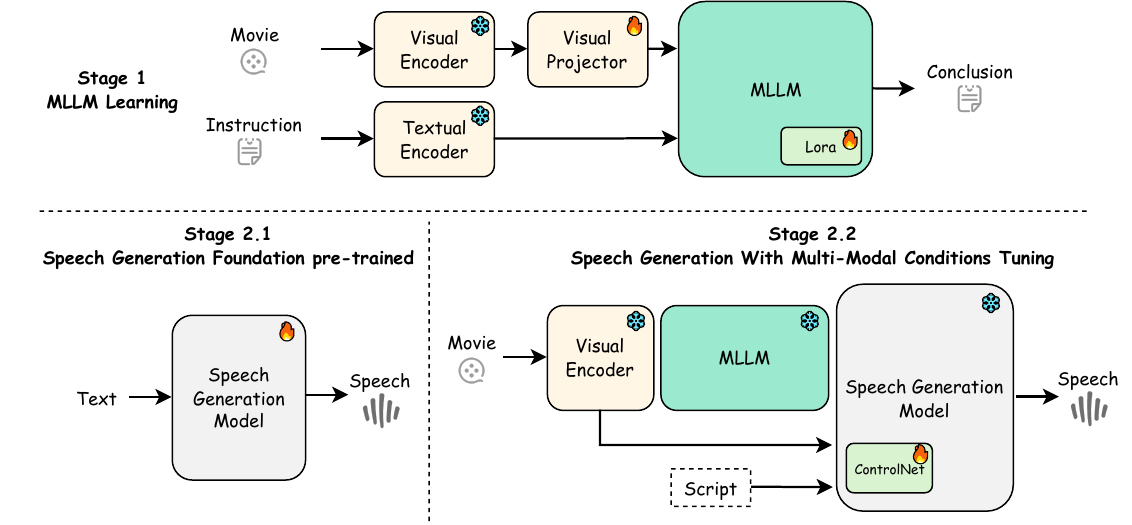}
    \caption{MM-MovieDubber pipeline with multi-stage, multi-modal training.}
    \label{fig:model_pipeline}
\end{figure*}

\section{Method}

\subsection{Overview}

Given a silent video clip \(V_l\), a corresponding subtitle \(T_v\), and the goal of generating a fully dubbed video, the proposed model (MM-MovieDubber) aims to produce speech \( \hat{S} \) that matches the video, ensures contextual and prosodic relevance, and maintains speech-video synchronization. The model can be formalized as follows:
\begin{equation}
\hat{S} = F_{dubber}^n(V_l, T_v)
\end{equation} 
MM-MovieDubber consists of two modeling stages: i) Multi-modal understanding through in-context learning. In this stage, a multi-modal large language model (mLLM) receives an instruction to generate an understanding conclusion \(C_{v}\) from video inputs, which can be formalized as:
\begin{equation}
C_{v} = F_{mm}^1(V_l, I_v)
\end{equation} 
\noindent where \(V_l\) represents the input video clip, and \(I_v\) represents the specified instruction. ii) Speech Generation Stage: This stage incorporates a conditional DiT based speech generator. The generative model combines the generated understanding conclusion, video clip and video related script to produce speech, which can be formalized as:
\begin{equation}
\hat{S} = F_{generator}^2(V_l, C_{v}, T_v)
\end{equation} 
\noindent where \(V_l\) represents the input video clip, \(C_{v}\) represents the understanding conclusion, and \(T_v\) represents the expected script. The speech \( \hat{S} \) is generated under conditioning.

\subsection{Stage 1: Multi-modal Large Language Model (mLLM) Learning}

Building upon the success of multi-modal large language models (mLLMs) like Visual-LLM \cite{liu2023visual,xu2025llavacotletvisionlanguage}, we employ multi-modal instruction tuning to achieve video understanding.
\begin{align}
    C_i &= F_{mm}(V_l, I_v) 
\end{align}
\noindent where \(V_l\) represents the input video clip, \(I_v\) represents the specific instruction, and \(F_{\text{mm}}\) represents the learning stage of the mLLM. Once the scene-related visual features are aligned with the textual scene features in the mLLM embedding space, the mLLM effectively encodes multi-modal information, which serves as a representation for understanding conclusion generation.

\begin{table*}[th]
\caption{Objective results on V2C-Animation benchmark. For the Dub 1.0 setting, we use the ground truth audio as reference audio, for the Dub 2.0 setting, we use the non-ground truth audio from the same speaker within the dataset as the reference audio which is more aligned with practical usage in dubbing.}
\centering
\resizebox{\linewidth}{!}{
    \begin{tabular}{c|c|ccccc|ccccc}
    \toprule
    Setting &  & \multicolumn{5}{c|}{Dub 1.0} & \multicolumn{5}{c}{Dub 2.0} \\ \midrule
    Methods & Visual & SPK-SIM (\%) $\uparrow$ & WER (\%) $\downarrow$ & EMO-SIM (\%) $\uparrow$ & MCD $\downarrow$ & MCD-SL $\downarrow$ & SPK-SIM (\%) $\uparrow$ & WER (\%) $\downarrow$ & EMO-SIM (\%) $\uparrow$ & MCD $\downarrow$ & MCD-SL $\downarrow$ \\ \midrule
    GT & - & 100 & 17.38 & 100 & 0.00 & 0.00 & 100 & 17.38 & 100 & 0.00 & 0.00 \\ \midrule
    F5-TTS \cite{chen2024f5ttsfairytalerfakesfluent} & \ding{55} & 89.30 & 24.41 & 76.78 & 8.32 & 8.32 & 83.11 & 24.83 & 64.91 & 10.86 & 10.87 \\ \midrule
    HPMDubbing \cite{congLearningDubMovies2023} & \ding{51} & 73.64 & 151.02 & 39.85 & 8.59 & 8.32 & 73.01 & 150.83 & 34.69 & 9.11 & 12.15 \\
    Speaker2Dub \cite{zhang2024from} & \ding{51} & 82.15 & 31.23 & 65.92 & 10.68 & 11.21 & 79.53 & 31.28 & 59.71 & 11.16 & 11.70 \\
    StyleDubber \cite{cong2024styledubbermultiscalestylelearning} & \ding{51} & 82.48 & 27.36 & 66.24 & 10.06 & 10.52 & 79.81 & 26.48 & 59.08 & 10.56 & 11.05 \\ \midrule
    Ours & \ding{51} & \textbf{89.74} & \textbf{22.52} & \textbf{78.88} & \textbf{6.98} & \textbf{6.99} & \textbf{83.30} & \textbf{24.71} & \textbf{64.93}& \textbf{8.80} & \textbf{8.80} \\ \bottomrule
    \end{tabular}
}
\label{table:v2c_benchmark}
\end{table*}

\begin{table*}[th]
\caption{Results on GRID benchmark with the same dub setting as the V2C-Animation benchmark.}
\centering
\resizebox{\linewidth}{!}{
    \begin{tabular}{c|c|cccccc|cccccc}
    \toprule
    Setting &  & \multicolumn{6}{c|}{Dub 1.0} & \multicolumn{6}{c}{Dub 2.0} \\ \midrule
    Methods & Visual & LSE-C $\uparrow$ & LSE-D $\downarrow$ & SPK-SIM (\%) $\uparrow$ & WER (\%) $\downarrow$ & MCD $\downarrow$ & MCD-SL $\downarrow$ & LSE-C $\uparrow$ & LSE-D $\downarrow$ & SPK-SIM (\%) $\uparrow$ & WER (\%) $\downarrow$ & MCD $\downarrow$ & MCD-SL $\downarrow$ \\ \midrule
    GT & - & 7.18 & 13.36 & 100 & 13.67 & 0.00 & 0.00 & 7.18 & 13.36 & 100 & 13.67 & 0.00 & 0.00 \\ \midrule
    F5-TTS \cite{chen2024f5ttsfairytalerfakesfluent} & \ding{55} & 5.51 & 14.70 & \textbf{96.51} & \textbf{11.94} & \textbf{4.23} & \textbf{4.24} & 5.10 & 14.71 & 94.45 & 16.75 & 4.89 & 4.90 \\ \midrule
    HPMDubbing \cite{congLearningDubMovies2023} & \ding{51} & \textbf{6.35} & 14.78 & 93.64 & 16.78 & 4.57 & 4.85 & \textbf{6.34} & 14.79 & 92.84 & 17.40 & 4.95 & 5.24 \\
    Speaker2Dub \cite{zhang2024from} & \ding{51} & 5.64 & 14.82 & 96.11 & 12.11 & 7.85 & 8.01 & 5.56 & 14.84 & 94.91 & 12.89 & 7.57 & 7.73 \\
    StyleDubber \cite{cong2024styledubbermultiscalestylelearning} & \ding{51} & 6.19 & 14.81 & 96.40 & 11.97 & 7.71 & 7.81 & 6.16 & 14.83 & \textbf{95.25} & \textbf{11.97} & 7.34 & 7.43 \\ \midrule
    Ours & \ding{51} & 4.87 & \textbf{14.63} & 95.73 & 14.71 & 4.48 & 4.49 & 4.46 & \textbf{14.63} & 94.71 & 16.08 & \textbf{4.73} & \textbf{4.74} \\ \bottomrule
    \end{tabular}
}
\label{table:grid_benchmark}
\end{table*}

\begin{figure*}[t!]
    \centering
    \includegraphics[scale=0.9]{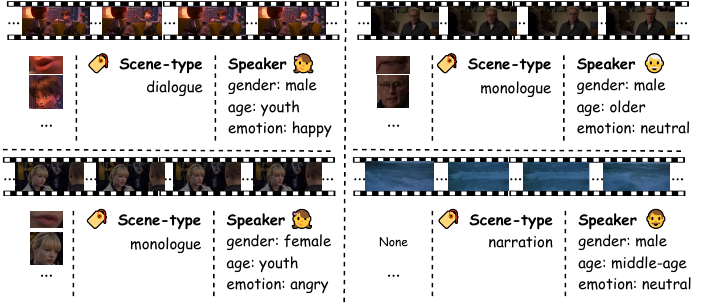}
    \caption{Proposed dataset with multi-type annotations, including annotation for lips, faces, scene-type, speaker gender, speaker age, voice emotion.}
    \label{fig:dataset}
\end{figure*}

\subsection{Stage 2: Mutli-Conditioned Speech Generation}

\subsubsection{Stage 2.1: Speech Generation Foundation Pre-Training}
In the second stage of MM-MovieDubber, the foundational speech generation model will first be trained. We use the same architecture as F5-TTS \cite{chen2024f5ttsfairytalerfakesfluent}, which employs a diffusion transformer (DiT) as the backbone and is trained to output a vector field $v_\tau$ with the conditional flow-matching objective $\mathcal{L}_{CFM}$ \cite{lipman2023flow}. 
The neural network’s training is guided by the conditional flow matching objective:
\begin{equation}
\mathcal{L}^{\text{CFM}}(\theta) = \mathbb{E}_{\tau, q(x_1), p(x|x_1)} \lVert u_\tau(x|x_1) - v_\tau(x; \theta) \rVert^2
\end{equation}
\noindent where $p_\tau$ is the probability path at time $\tau$, $u_\tau$ is the designated vector field for $p_\tau$, $x_1$ symbolizes the random variable corresponding to the training data, and $q$ is the distribution of the training data. 

\subsubsection{Stage 2.2: Speech Generation With Multi-Modal Conditions Tuning}

Next, in the ControlNet-transformer tuning stage, the video frames, along with an instruction
are fed as inputs to the mLLM model. The sequence of video features and the video understanding conclusion are then combined and passed into the speech generation model. In this context, the provided conclusion helps guide the video-to-speech generation process, as shown in stage 2.2 of Figure \ref{fig:model_pipeline}. The proposed speech generation model takes as input the script, silent video, video understanding conclusion, and an optional reference speech, and generates video-aligned speech context sequences, which can be described as:
\begin{equation}
\hat{S} = F_{generator}(V_l, C_v \{C_1, \ldots, C_n\}_{n=4}, T_v)
\end{equation} 
\noindent where $C_v$ is the combination of \(\{C_1, \ldots, C_n\}, n=4\), which represents the scene type condition $C_s$, speaker gender condition $C_g$, speaker age condition $C_a$, and speaker emotion condition $C_e$. These conditions are combined and encoded by an encoder \cite{RaffelSRLNMZLL20}. The $V_l$ represents the visual features derived from the input video frames, which are encoded by CLIP \cite{RadfordKHRGASAM21}.
We implement a cross-attention mechanism that facilitates the integration of understanding conclusion features $C_v$ and visual features $V_l$. 
Additionally, \(T_v\) represents the embedded script. Furthermore, we added a duration loss \(\mathcal{L}_{dur}\) to constrain the duration consistency, which can be described as:
\begin{equation}
\mathcal{L}_{\text{dur}} = \ell\big( f(V_l, C_l), \, dur \big) ,
\end{equation} 
The final loss function is constructed as follows:
\begin{equation}
    \resizebox{.9\linewidth}{!}{$
    \mathcal{L}_{\text{g}} = \mathbb{E}_{\tau, q(x_1), p(x|x_1)} \lVert u_\tau(x|x_1, v_l, c_v, t_v) - v_\tau(x; \theta) \rVert^2 + \mathcal{L}_{\text{dur}}
$}
\end{equation}
In the training stage, the visual condition $V_l$, video understanding conclusion condition $C_v$, and video script condition $T_v$ are each set to $\phi$ with a 5 \% probability.
Extending classifier-free guidance from the script condition to visual input and visual understanding enhances both conditional control precision and speech quality. The guidance scales, $\lambda_V$, $\lambda_C$, and $\lambda_T$, correspond to the video clip, video conclusion, and video-related script, respectively, and measure the alignment between the sampling results and conditions. Inspired by \cite{jiang2024dive}, during inference, the modified velocity estimate is as follows:
\begin{equation}
\resizebox{.68\linewidth}{!}{$
\begin{aligned}
g_{\theta}' & = g_{\theta}(x_\tau, \phi, \phi, \phi) \\ 
 & + \lambda_V \cdot (v_0(x_\tau, c_v, c_c, c_t) - v_0(x_\tau, \phi, c_c, c_t)) \\ 
 & + \lambda_C \cdot (v_0(x_\tau, \phi, c_c, c_t) - v_0(x_\tau, \phi, \phi, c_t)) \\ 
 & + \lambda_T \cdot (v_0(x_\tau, \phi, \phi, c_t) - v_0(x_\tau, \phi, \phi, \phi)) \\
\end{aligned}
$}
\end{equation}

\section{Experiments}

\subsection{Datasets}

\noindent\textbf{Emilia} is a comprehensive multilingual speech generation dataset containing a total of 101,654 hours of speech data across six languages \cite{he2024emiliaextensivemultilingualdiverse}. 
The English portion of this dataset, comprising approximately 46,800 hours, is utilized to train our foundational text-to-speech (TTS) model.

\noindent\textbf{V2C-Animation} is a specialized dataset designed for animated movie dubbing, consisting of 10,217 clips from 26 films with synchronized text, audio, and video \cite{chen2021v2cvisualvoicecloning}. 
The dataset is partitioned into 60\% for training, 10\% for validation, and 30\% for testing.

\noindent\textbf{GRID} is a dubbing benchmark for multi-speaker dubbing \cite{10.1121/1.2229005}. The whole dataset has 33 speakers, each with 1000 short English samples. All participants are recorded in studio with unified background. The number of train and test data are 32,670 and 3280, respectively.

\noindent\textbf{Movie Dubbing Dataset.}
We have built a 7.2-hour multi-modal movie dubbing dataset with multi-type annotations
to enhance movie understanding and improve dubbing quality.
As shown in Figure \ref{fig:dataset}, the conclusion provides a high-level overview of the entire scene, while other labels describe the scene type, speaker, and speech in the video. The dataset is partitioned as follows: 60\% for training, 10\% for validation, and 30\% for testing.

\subsection{Implementation Details}
To improve training stability, multi-modal speech generation capability is progressively incorporated throughout the training process using a multi-stage approach. Video frames are sampled at 25 FPS, and all speech is resampled to 22 kHz. The foundational speech generation model is trained on 8 NVIDIA A800 80G GPUs for about one week using Emilia-EN \cite{he2024emiliaextensivemultilingualdiverse}. The mLLM learning stage and the speech generation tuning stage are implemented on a single NVIDIA A800 80G GPU with the proposed movie-dubbing dataset.

\section{Evaluation}
We evaluate using both objective and subjective metrics. To assess pronunciation accuracy, we use Word Error Rate (WER) with Whisper-V3\cite{radford2022robustspeechrecognitionlargescale} as the ASR model. Timbre consistency is evaluated with speaker encoder cosine similarity (SPK-SIM) \cite{cong2024styledubbermultiscalestylelearning}. We also calculate mel cepstral distortion dynamic time warping (MCD) and speech length variance (MCD-SL) \cite{battenberg2020locationrelativeattentionmechanismsrobust} for spectral and length differences. Emotion similarity is assessed using a speech emotion recognition model \cite{ye2023temporal}. For alignment with video, we use Lip Sync Error Distance (LSE-D) and Lip Sync Error Confidence (LSE-C) metrics on the Grid benchmark, based on the pre-trained SyncNet model \cite{chung2017out}. For subjective evaluation, we conduct human evaluations of the Mean Opinion Score (MOS) for naturalness (NMOS) and similarity (SMOS), rated on a 1-to-5 scale with 95\% confidence intervals. Following \cite{zhang2024from}, participants evaluate the dubbing quality of 30 randomly selected audio samples from each test set.

\subsection{Results}

We compare our approach with a recent SOTA TTS model, F5-TTS \cite{chen2024f5ttsfairytalerfakesfluent}, and three recent SOTA video dubbing models: HPMDubbing \cite{congLearningDubMovies2023}, StyleDubber \cite{cong2024styledubbermultiscalestylelearning}, and Speaker2Dubber \cite{zhang2024from}.
Tables \ref{table:v2c_benchmark}, \ref{table:grid_benchmark}, \ref{table:zeroshot_test}, and \ref{table:initial_resoning} present our objective evaluation results. 
In the subjective evaluation, we randomly select $30$ samples from the generated dubbing of each dub setting for human assessment, and the results are shown in Table \ref{table:mos_benchmark}.

\subsubsection{Comparison with SOTA}

\noindent\textbf{Results on V2C-Animation benchmark.}
As shown in Table \ref{table:v2c_benchmark}, in the comparison with the state-of-the-art models\cite{congLearningDubMovies2023,cong2024styledubbermultiscalestylelearning,zhang2024from}, our model achieves improvements across evaluation metrics in the same setting\cite{zhang2024from}. In details, the SPK-SIM increased from 79.81\% to 83.30\%, EMO-SIM improved from 59.71\% to 64.93\%, MCD decreased from 9.11 to 8.80, and WER reduced from 26.48\% to 24.71\%.

\noindent\textbf{Results on GRID benchmark.}
As shown in Table \ref{table:grid_benchmark}, our model achieves the best lip-sync performance on the GRID benchmark with the same evaluation of the state-of-the-art models\cite{zhang2024from}, which decreased from 14.79 to 14.63. And the MCD decreased from 4.95 to 4.73. 
In addition, our method also achieves competitive results in WER, slightly lower than the best fine-tuned F5-TTS model, Speaker2Dub and StyleDubber. However, these models have WER results (11.94\%, 12.11\% and 11.97\%) that exceed the ground truth WER result (13.67\%), suggesting that the intelligibility has reached an acceptable range for humans.

\begin{table}[]
\centering
\caption{Subjective evaluation on V2C-Animation and GRID benchmarks.}
\resizebox{0.9\linewidth}{!}{
    \begin{tabular}{c|cc|cc}
    \toprule
    Dataset & \multicolumn{2}{c|}{V2C-Animation} & \multicolumn{2}{c}{GRID} \\ \midrule
    Methods & NMOS & SMOS & NMOS & SMOS \\ \midrule
    GT & 4.98±0.01 & - & 4.99±0.01 & - \\ \midrule
    F5-TTS \cite{chen2024f5ttsfairytalerfakesfluent} & 4.20±0.68 & 3.83±0.63 & \textbf{4.43±0.03} & \textbf{3.32±0.05}  \\ \midrule
    HPMDubbing \cite{congLearningDubMovies2023} & 1.04±0.01 & 1.02±0.01 & 3.50±0.10 & 2.77±0.12 \\
    Speaker2Dubber \cite{zhang2024from} & 2.93±0.21 & 2.58±0.19 & 4.04±0.07 & 3.00±0.10 \\
    StyleDubber \cite{cong2024styledubbermultiscalestylelearning} & 2.68±0.21 & 2.39±0.21 & 4.01±0.03 & 3.06±0.07 \\ \midrule
    Ours & \textbf{4.37±0.35} & \textbf{3.91±0.45} & 4.33±0.07 & 3.14±0.08 \\ \bottomrule
    \end{tabular}
}
\label{table:mos_benchmark}
\end{table}

\begin{table}[]
\centering
\caption{Results on zero-shot test, which use unseen speaker as reference audio.}
\resizebox{\linewidth}{!}{
    \begin{tabular}{c|cccccc}
    \toprule
    Setting & \multicolumn{5}{c}{Dubbing Setting 3.0} \\ \midrule
    Methods & LSE-C $\uparrow$ & LSE-D $\downarrow$ & SPK-SIM $\uparrow$ (\%) & WER $\downarrow$ (\%) & MOS $\uparrow$ \\ \midrule
    HPMDubbing \cite{congLearningDubMovies2023} & 1.72 & \textbf{11.74} & 68.14 & 126.85 & 1.29±0.60  \\
    Speaker2Dub \cite{zhang2024from} & 2.21 & 12.67 & 76.10 & 16.57 & 3.38±0.14 \\
    StyleDubber \cite{cong2024styledubbermultiscalestylelearning} & 2.15 & 12.76 & 78.30 & 19.07 & 3.30±0.15 \\ \midrule
    Ours & \textbf{2.21} & 12.59 & \textbf{83.55} & \textbf{15.49} & \textbf{4.12±0.16} \\ \midrule
    \end{tabular}
}
\label{table:zeroshot_test}
\end{table}

\begin{table}[]
\caption{Results on proposed benchmark, which use silence video and script as inputs.}
\centering
\resizebox{\linewidth}{!}{
    \begin{tabular}{c|ccccc}
    \toprule
    Setting & \multicolumn{5}{c}{towards fined-grained movie dubbing test} \\ \midrule
    Methods & LSE-C $\uparrow$ & LSE-D $\downarrow$ & SPK-SIM (\%) $\uparrow$ & WER $\downarrow$ (\%) & MOS \\ \midrule
    HPMDubbing~\cite{congLearningDubMovies2023} & 1.46 & \textbf{11.46} & 61.06 & 199.40 & 1.11±0.06 \\
    Speaker2Dub~\cite{zhang2024from} & 1.68 & 12.00 & 61.73 & 84.42 & 1.54±0.22 \\
    StyleDubber~\cite{cong2024styledubbermultiscalestylelearning} & \textbf{2.35} & 13.06 & 64.03 & 52.69 & 2.76±0.16 \\ \midrule
    Ours & 1.84 & 11.88 & \textbf{70.82} & \textbf{27.68} & \textbf{4.01±0.04} \\ \midrule
    \end{tabular}
}
\label{table:initial_resoning}
\end{table}

\begin{table}[ht]
\centering
\caption{Results of ablation study on the proposed dataset.}
\resizebox{\linewidth}{!}{
    \begin{tabular}{c|cccccc}
    \toprule
     & LSE-C $\uparrow$ & LSE-D $\downarrow$ & SPK-SIM (\%) $\uparrow$ & EMO-SIM (\%) $\uparrow$ & MCD $\downarrow$ & MCD-SL $\downarrow$  \\ \midrule
    w/o clip & 1.99 & 12.73 & 82.63 & 63.24 & 8.85 & 8.86 \\
    w/o dur & \textbf{2.06} & 12.81 & 82.71 & 64.32 & 8.82 & 8.83 \\ \midrule
    w/o conclusion & 1.89 & 12.66 & 82.59 & 63.18 & 8.81 & 8.82 \\
    proposed & 2.01 & \textbf{12.61} & \textbf{82.99} & \textbf{64.74} & \textbf{8.76} & \textbf{8.77} \\ \bottomrule
    \end{tabular}
}
\label{table:aba_res}
\end{table}

\noindent\textbf{Results on Speaker Zero-shot test.}
This setting uses the audio of unseen speakers as reference audio to measure the generalizability of the dubbing model\cite{zhang2024from}. Here, we use the audio from GRID as reference audio to measure V2C. We compare LSE-C/D, SPK-SIM, and WER, along with subjective evaluations in the same evaluation setting with the state-of-the-art models\cite{zhang2024from}. As shown in Table \ref{table:zeroshot_test}, our method outperforms the StyleDubber and Speaker2Dub in both SPK-SIM and WER. In details, the SIP-SIM improved from 78.30\% to 83.55\%, the WER decreased from 16.57\% to 15.49\%.
Furthermore, the proposed method still maintains competitive performance in audio-visual synchronization (see LSE-C and LSE-D), slightly lower than HPMDubbing.

\noindent\textbf{Results on towards fined-grained movie dubbing test.}
In order to better fit the real-world usage scenarios, we also conducted a test on the proposed dataset, which we call the towards fined-grained movie dubbing test. This setting just uses the silent video and related script as inputs (no reference speech) to measure the performance of the dubbing model for different dubbing types(Dialogue, Narration and Monologue) and fine-grained attributes(gender, age and emotion). The results are shown on Table ~\ref{table:initial_resoning}. Our method outperforms the SOTA dubbing methods (StyleDubber and Speaker2Dub) on SPK-SIM, EMO-SIM and WER. In details, the SPK-SIM improved from 64.03\% to 70.83\%, WER decreaded from 52.69\% to 27.68\%. Furthermore, the proposed method still maintains the competitive performance in audio-visual synchronization (see LSE-D), 11.88\%, which slightly lower than HPMDubbing(11.46\%).

\noindent\textbf{Ablation Studies.}
The ablation results in Table \ref{table:aba_res} indicate that each condition plays a role in overall performance. Removing the video clip control causes all metrics to drop significantly, showing its importance for speech-video alignment. Adding the video understanding conclusion control improves SPK-SIM and EMO-SIM. Finally, removing the duration predictor leads to the largest drop in LSE-D performance, emphasizing that duration-level consistency learning is essential for synchronizing speech and video.

\section{Conclusion}
In this paper, we propose a multi-stage, multi-modal generative framework for movie dubbing. Additionally, we have developed a movie dubbing dataset with multi-type annotations to enhance movie understanding and improve dubbing quality. In the future, we plan to incorporate reasoning ability into our movie dubbing model to improve its understanding and reasoning skills, ultimately producing higher-quality dubbing.





\bibliographystyle{IEEEtran}
\bibliography{mybib}

\end{document}